\documentclass[sn-mathphys-num]{sn-jnl}

\usepackage{graphicx}%
\usepackage{multirow}%
\usepackage{amsmath,amssymb,amsfonts}%
\usepackage{amsthm}%
\usepackage{mathrsfs}%
\usepackage[title]{appendix}%
\usepackage[table,x11names]{xcolor} 

\usepackage{textcomp}%
\usepackage{manyfoot}%
\usepackage{booktabs}%
\usepackage{algorithm}%
\usepackage{algorithmicx}%
\usepackage{algpseudocode}%
\usepackage{listings}%
\usepackage{tabularx}%
\usepackage{booktabs}
\usepackage{hanging}

\title{Equity in the Use of ChatGPT for the Classroom: A Comparison of the Accuracy and Precision of ChatGPT 3.5 vs. ChatGPT4 with Respect to Statistics and Data Science Exams}

\author*[1]{\fnm{Monnie} \sur{McGee}}\email{mmcgee@smu.edu}
\equalcont{These authors contributed equally to this work.}

\author*[1]{\fnm{Bivin} \sur{Sadler}}\email{bsadler@smu.edu}
\equalcont{These authors contributed equally to this work.}

\affil*[1]{\orgdiv{Department of Statistics and Data Science}, \orgname{Southern Methodist University}, \orgaddress{\street{6425 Boaz Lane}, \city{Dallas}, \postcode{75205}, \state{TX}, \country{USA}}}
\vspace{-5mm}

\begin{document}
\maketitle

\vspace{-1cm}
{\bf Abstract:} A college education historically has been seen as  method of moving upward with regards to income brackets and social status. Indeed, many colleges recognize this connection and seek to enroll talented low income students. While these students might have their education, books, room, and board paid; there are other items that they might be expected to use that are not part of most college scholarship packages. One of those items that has recently surfaced is access to generative AI platforms. The most popular of these platforms is ChatGPT, and it has a paid version (ChatGPT4) and a free version (ChatGPT3.5). We seek to explore differences in the free and paid versions in the context of homework questions and data analyses as might be seen in a typical introductory statistics course. We determine the extent to which students who cannot afford newer and faster versions of generative AI programs would be disadvantaged in terms of writing such projects and learning these methods. 
\vspace{5mm}

\noindent{\bf Authors' Note:} This article was originally submitted to a journal in May of 2024 and, after waiting for reviews for 6 months, was rejected. While the content is obsolete because ChatGPT3.5 is no longer available, we hope that this paper will serve to set precedent and to help others performing a similar analysis. All code and data sets have been deposited in the second author's GitHub repository: https://github.com/BivinSadler/generativeAIEquityStudy.

\vspace{5mm}
\noindent{\bf Keywords:} Introductory statistics, generative AI, data analysis, knowledge gap, digital divide, McNemar's test, ordinal logistic regression

\section{Introduction}\label{sec1}

The association of social mobility with a college education has been studied since the early 1950’s \cite{reber}. Although there are some indications that a college education is not as effective as it once was in helping graduates climb the social ladder \cite{tough}, it is still the most reliable way of doing so. US News \& World Report updated its rankings in 2023 to include social mobility \cite{brooks}, and many institutions of higher education are paying more attention to recruitment of first-generation college students and talented students from disadvantaged backgrounds. With the inclusion of such students in the typical college class comes some important considerations. For example, a student from difficult financial circumstances with an academic background to match the profile of any student an elite institution will have more difficulty paying for textbooks, a laptop, a smartphone, and other items that are almost essential to current college life \cite{tough}. 

As of November 2022, one such item that students from advantaged backgrounds will have access to that those from lower income brackets will not is ChatGPT4 \cite{gpt4}. It currently costs \$20 per month for a subscription and has been called a ``significant leap forward''  compared to  ChatGPT3.5 \cite{gpt3}, which is free \cite{gupta}. While use of generative AI is prohibited in some college classrooms, this is hard to police, and many students use it regardless of classroom restrictions \cite{terry}. When generative AI is allowed, there is a wide array of platforms from which students can choose. The platforms include Anthropic's Claude \cite{claude}, Google Gemini \cite{gemini}, Microsoft Co-Pilot \cite{copilot}, Pi \cite{pi}, ChatGPT3.5 \cite{gpt3}, and ChatGPT4 \cite{gpt4}. As of the writing of this article, Pi and ChatGPT3.5 are free for use. Claude and Gemini have both free and ``pro'' versions, where the pro version promises greater functionality and accuracy for the price of approximately \$20 per month, much like ChatGPT4. Copilot is currently free with a license for Office 365 \cite{oit}. 

Regardless of the platform used, some students will be able to pay for newer, faster, more equipped versions, while others will need to use the free version. Many believe that generative AI platforms have the potential to revolutionize education and shrink or even close the gap between the haves and have nots with respect to education \cite{ellis}. This belief assumes that each student has access to the same version of a  generative AI platform. A significant difference in performance between the free and paid versions of generative AI platforms would, ironically, widen the knowledge gap even further. 

Admittedly, lack of financial resources is not the only reason someone would not want to purchase a paid version of any generative AI platform. Some students or faculty simply do not want to give their credit card information to an entity that is untested. Some individuals might not think they will use it enough to justify the cost. Others might not want the extra complexity and services that come with a monthly subscription. In addition, there are other issues of equity and access besides the ability or desire to pay for course materials or study aids. One accessibility issue involves the visual nature of these platforms. All of them involve typing text and reading the output. Both typing and reading might be barriers to some. To overcome these barriers, some platforms have text-to-speech capability. Pi, Gemini, and ChatGPT all support the use of screen readers and speech-to-text software. We did not test either screen readers or speech-to-text thoroughly; however, both ChatGPT3.5 and ChatGPT4 seemed to cooperate well with screen readers and speech-to-text software in our limited testings.

The purpose of this research is to determine differences in quality and accuracy between ChatGPT3.5 and ChatGPT4 for typical questions that might be asked in the context of a statistics and data science course. Specifically, we examined correctness with regards to answering questions on four different exams: the Arkansas Council of Teachers of Mathematics (ACTM) exam \cite{ACTMExam}, Comprehensive Assessment of Outcomes in Statistics (CAOS) \cite{caos},  2011 AP statistics exam, and an exam given in a first-year graduate statistical methods course for PhD students in statistics, biostatistics, and data science. We collected exam questions from these sources and entered the questions, one at a time, into ChatGPT3.5 and ChatGPT4. Questions were of different types, such as multiple  choice and free response. Some questions included visualization of data, and some did not. We saved the answers generated by each platform and graded them independently. For each type of question and exam, we compared the scores for the two platforms. A comparison of the scores shows the extent to which students who use free versions of generative AI programs would be disadvantaged in terms of writing such projects and learning these methods.

It should be noted that, while we examine the quality of results from the platforms using nationally normed exams, we do not condone students' use of generative AI to take exams. The exams used in this study are a convenient method to measure the quality of answers students would receive if using generative AI to review for these exams or to study for a statistics course. 

\section{Literature Review}\label{sec:lit}

Much of the existing literature has explored the ethics of potential general uses for generative AI in the classroom, as well as some of the risks \cite{gill2024}.  Ethical issues, reliability, and robustness have been identified as considerations for the use of AI generated content (AIGC) in engineering management \cite{yugong}. Specifically, a rigorous system of checks and balances on AIGC is recommended to avoid potential safety concerns. Such a system does not yet exist. A review paper on the use of AI in education gives more details \cite{baidoo}.

Most comparisons of ChatGPT3.5 to ChatGPT4 have been done in the context of performance on standardized exams. For example, ChatGPT3.5 performs worse than ChatGPT4 on the American Academy of Opthomology exam \cite{taloni}. The performance of ChatGPT3.5 vs.~ChatGPT4 has also been compared using the ACT, SAT, and other college entrance exams, in addition to many other typical student assignments \cite{openai2024gpt4}.  ChatGPT3.5 received a score of 68\% on the Bar Exam \cite{chat35bar}, while a different study showed ChatGPT4 passed the bar with a score much better than the average for humans \cite{chat4bar}. Physicians have tested the platforms' ability to diagnose certain ailments \cite{cma}.  In all of the studies of this nature done thus far, ChatGPT4 has been the clear winner.  

Contributions to the use of generative AI in the statistics and data science classrooms have included blogs \cite{virginia} and articles \cite{differentEllis}.  
To our knowledge, comparisons of the performance of ChatGPT3.5 and ChatGPT4 with regard to student learning outcomes in statistics and data science courses have not been published; however, several articles on the use of generative AI as a tool to improve student student learning, mostly in pre-post designs, have been published. For example, generative AI has been examined as a tutor for second-language learners \cite{huang23} or to improve student writing skills \cite{Hidayatullah_2024}. Another study used a SWOT (strength, weaknesses, opportunities, threats) analysis for 58 students in a linguistics course who were using generative AI to learn basic machine learning techniques for natural language processing (NLP) \cite{tay2023}. A recent study of responses to questions on computer science exams examined accuracy of responses from ChatGPT3.5 on core subjects in computer science,  the graduate aptitude test in engineering, and programming using LeetCode \cite{joshi}. This study did not compare performance between ChatGPT3.5 and ChatGPT4, but rather examined performance of  ChatGPT3.5. The authors found an overall average score of 58.9\% for the subjects tested, with best performance on coding and worst performance on data structures. A comparison in the content of lesson plans generated by ChatGPT and Gemini for seventh grade mathematics, science, English, and social studies examined 18 different lesson plans according to best practices in learning theory \cite{Baytak_2024}. Note than in both the computer science and lesson plan studies, free versions of generative AI platforms were used for comparison. The lack of comparison with ChatGPT4 might in itself might be revealing, as both articles were published after ChatGPT4 was available. Perhaps the fact that some generative AI platforms require a monthly fee for access is also a barrier for faculty research, as well as a barrier for student use.

\section{Methods}\label{methods}

We selected four different exams representing various levels of statistics knowledge to use in this study. Two of them were nationally normed exams, the Comprehensive Assessment of Outcomes in Statistics (CAOS) \cite{caos} and the AP statistics exam. The CAOS exam was originally written to gauge student learning outcomes in college-level introductory statistics courses. The AP exam is meant to replace a college level statistics course; therefore, it can be considered as a level above the CAOS exam. We also a standardized  test designed by the Arkansas Council of Teachers of Mathematics (ACTM) \cite{ACTMExam}, which is meant to quantify understanding of statistical concepts at the high school level. Finally, we used questions from a homemade exam given in a first-year graduate course in statistical methods for PhD students in statistics, biostatistics, and data science. Questions from this last exam are meant to represent assessment of statistical knowledge at the graduate level. Regardless of the level of exam, all exams consisted of two main types of questions: multiple choice and free response. Each type of question has questions where an image was displayed or not. Therefore, there were four types of questions: multiple choice without an image (MC, $n=43$), multiple choice with an image (MCI, $n=22$), free response without an image (FR, $n=20$) and free response with an image (FRI, $n=8$). Examples of each type of question are given in Appendix \ref{secA1}. 

The procedure was to copy and paste each question from a given exam into ChatGPT 3.5 or 4 one at a time. If ChatGPT gave a partially correct response, we did not prompt it for more information. We did this because we believe that is what a student using ChatGPT would do - the student would take the first answer given by ChatGPT rather than ask further questions to obtain a complete answer. We then scored the answer from ChatGPT for MC and MCI questions as either 3 for correct or 0 for incorrect. Correct MC questions were assigned a value of 3 to be comparable to the scale of the free response questions, which were graded on a scale of 0 to 4. The scale is given in Table \ref{tab:rubric}. In this rubric, a response of 3 is a correct, acceptable response.

\begin{table}[ht]
\caption{Scoring rubric for free response and free response questions with images.}\label{tab:rubric}
\begin{tabular}{|p{1cm}|p{7cm}|}
\hline
Score & Reasoning \\
\hline
0 & Completely Incorrect: All responses are either incorrect or missing. \\
1 & Mostly Incorrect: Most of the responses are incorrect although some are relevant and correct. \\
2 & Somewhat Incorrect: At least some but not most of the responses are incorrect or irrelevant.\\
3 & Acceptable: This is the minimal viable response. There would be no points deducted.\\
4 & Good: The response is correct, well written and insightful. \\
\hline
\end{tabular}
\end{table}

Once the scores were tabulated, we analyzed the results using descriptive statistics and visualizations. Differences in mean scores between ChatGPT3.5 and ChatGPT4 were examined using either McNemar's test \cite{McNemar1947} or ordinal logistic regression \cite{agresti}, depending on the question type. In subsections \ref{sec:mcnemar} and \ref{sec:olr}, we explain each of these analysis methods and justify the use of them for various question types from our study.

\subsection{McNemar's Test}\label{sec:mcnemar}

McNemar's test is a $\chi^2$ test designed for examining the relationship between two variables (ChatGPT3.5 and ChatGPT4, in this case) with two categories each where the units are paired. The multiple choice questions fit this scenario. The units are the questions, and each question was scored as either correct or incorrect. Because we entered the same questions into each version of ChatGPT, the outcomes are paired. McNemar's test examines the discrepancy between discordant versus concordant pairs. A discordant pair is one in which one the platforms provide different outcomes (ie. one of the platforms provides a correct solution while the other does not). A question generates a concordant pair when both platforms provide the same outcome (either both answer that question correctly or both incorrectly.)

The null and alternative hypotheses of McNemar's test are:

\vspace{3mm}
\begin{hangparas}{.25in}{1}
$H_0:$ The probability of ChatGPT4 getting a question right and ChatGPT3.5 getting it wrong is the same as ChatGPT4 getting a question wrong and ChatGPT3.5 getting it right.

$H_A:$ The probability of 4 getting a question right and 3.5 getting it wrong is different than the probability of 4 getting a question wrong and 3.5 getting it right.
\end{hangparas}
\vspace{3mm}

Table \ref{tab:ContingencyTable} displays a contingency table containing concordant and discordant pairs.  Within the table, the letter a represents the number of questions that both ChatGPT3.5 and ChatGPT4 answered correctly and d represents the number of questions they both answered incorrectly. Therefore,  a and d represent the number of concordant pairs.  The  letter b represents the number of questions that ChatGPT3.5 answered correctly, but 4.5 did not. The letter c represents the number of questions that ChatGPT4 answered correctly, but ChatGPT3.5 answered incorrectly.  Thus, b and c represent the number of discordant pairs. 

\begin{table}[h]
\centering
\caption{A Contingency Table}
\label{tab:ContingencyTable}
\begin{tabular}{lcc}
\toprule
 & \textbf{ChatGPT4 Correct} & \textbf{ChatGPT4 Incorrect} \\
\midrule
\textbf{ChatGPT3.5 Correct}   & a & b \\
\textbf{ChatGPT3.5 Incorrect} & c & d \\
\bottomrule
\end{tabular}
\end{table}

The test statistic for McNemar's test is calculated as:

\[
\chi^2 = \frac{(b - c)^2}{b + c}
\]

where:
\begin{align*}
&b = \text{Number of pairs where ChatGPT3.5 outperforms ChatGPT4} \\
&c = \text{Number of pairs where ChatGPT4 outperforms ChatGPT3.5.}
\end{align*}

To determine whether to reject the null hypothesis, we calculate a p-value based on the $\chi^2$  distribution with one degree of freedom. If the p-value is small, we reject the null hypothesis. Evidence as to which platform has a higher probability of being correct is determined by the sample proportions of correct responses for only the discordant pairs for each platform. 

\subsection{Ordinal Logistic Regression}\label{sec:olr}

Ordinal logistic regression is an extension of binary logistic regression. It is used to model the relationship between an ordinal response variable and one or more predictor variables. The predictor variables can be either categorical or quantitative. It is suitable when the response variable has ordered categories but the intervals between categories are not assumed to be equal. 

The model can be expressed as:

\[
\log\left(\frac{P(Y \leq j)}{P(Y > j)}\right) = \alpha_j - \mathbf{X}\beta
\]

\noindent where $P(Y \leq j)$ is the cumulative probability of the response $Y$ being less than or equal to category $j$, $\frac{P(Y \leq j)}{P(Y > j)}$ are known as the odds, $\alpha_j$ are the threshold parameters for each category $j$, $\mathbf{X}$ is a matrix of predictor variables, $\beta$ is a vector of coefficients associated with the predictor variables.

With logistic regression models, the odds ratio of an event is often calculated as part of the output. The odds represent the likelihood of an event occurring compared to the likelihood of it not occurring. In ordinal logistic regression, the odds ratio compares the odds of the response being in one category versus the odds of it being in a lower category. 

To estimate the odds ratio from the parameter estimates ($\beta$), the coefficient associated with the predictor variable is back transformed using the exponential function. For example, if $\beta_1$ is the coefficient for a predictor variable $X_1$, the odds ratio associated with a one-unit increase in $X_1$ is $e^{\beta_1}$. The mathematical relationship is shown in Equation \ref{eq:odds}.
\begin{equation}
e^{\beta_1}=\frac{P(Y \leq j \,|\, X_1 + 1)}{P(Y > j \,|\, X_1 + 1)}/{\frac{P(Y \leq j \,|\, X_1)}{P(Y > j \,|\, X_1)}} 
\label{eq:odds}
\end{equation}

 This study employs ordinal logistic regression to evaluate the relationship between a set of predictors and an ordinal response variable, which has ordered categories but unequal intervals. To accommodate the paired nature of the data, the response variable was defined as the difference between scores for the each question between ChatGPT4 and ChatGPT3.5. The scores for each question were given using the rubric in Table \ref{tab:rubric}. The differences ranged from scores $<0$ (indicating ChatGPT3.5 received a higher score than ChatGPT4) to 4 (ChatGPT4 was completely correct and ChatGPT3.5 was completely incorrect). We translated the response variable, indexed by \( j \) into the following comparative evaluations of ChatGPT4 and ChatGPT3.5:
 
\begin{itemize}
    \item[]\textbf{Much:} ChatGPT4 was ``Much better" than ChatGPT3.5.
    \item[] \textbf{Somewhat}: ChatGPT4 was ``Somewhat better" than ChatGPT3.5.
    \item[] \textbf{Same}: Both versions performed ``the same."
    \item[] \textbf{Worse}: ChatGPT4 performed ``Worse" than ChatGPT3.5.
\end{itemize}

The ordinal logistic regression model is formulated as:
\[
\log \left( \frac{P(Y \leq j)}{P(Y > j)} \right) = \alpha_j + \beta_1 \text{FR} + \beta_2 \text{Image}
\]
where \( \alpha_j \) are thresholds (intercepts) that define the boundaries between categories, and \( \beta_1 \) and \( \beta_2 \) are coefficients for the predictors:
\begin{itemize}
    \item[] \textbf{FR}: Indicator of Free Response (1) versus Multiple Choice (0) questions.
    \item[] \textbf{Image}: Presence (1) or absence (0) of an image in the question.
\end{itemize}

\textbf{Interpretation of Exponentiated Coefficients:} The exponentiated coefficients, \( e^{\beta_1} \) and \( e^{\beta_2} \), represent the multiplicative change of the odds of ChatGPT4 providing a higher quality response than ChatGPT3.5 for different levels of the Type question (FR or MC) and if the question has an image (0 or 1), respectively.  This multiplicative change is also known as the odds ratio. 
\begin{itemize}
    \item[] \( e^{\beta_1} \): The multiplicative change of the odds (the odds ratio) of ChatGPT4 providing a higher quality response than ChatGPT3.5 for Free Response questions compared to those that are Multiple Choice, holding the presence of an image constant.
    \vspace{2mm}
    
    \item[] \( e^{\beta_2} \): The multiplicative change of the odds (the odds ratio) of ChatGPT4 providing a higher quality response than ChatGPT3.5 for question with images compared to those that do not have images, holding the question type constant.
\end{itemize}

Often, the probability of being in a specific level is more useful than the odds.  The exact probability of ChatGPT4 being rated in each category for a given question can be calculated as:
\begin{equation}
P(Y = j) = \frac{e^{\alpha_j + \beta_1 \text{FR} + \beta_2 \text{Image}}}{1 + e^{\alpha_j + \beta_1 \text{FR} + \beta_2 \text{Image}}} - \frac{e^{\alpha_{j-1} + \beta_1 \text{FR} + \beta_2 \text{Image}}}{1 + e^{\alpha_{j-1} + \beta_1 \text{FR} + \beta_2 \text{Image}}}\label{eq:prob}
\end{equation}

In Section \ref{sec:results}, we first examine the accuracy for ChatGPT3.5 and ChatGPT4 for all questions, regardless of question type. Next, we compare the accuracy of the platforms for multiple choice questions, free response questions, questions with images, and questions without images. Finally, in Section \ref{sec:olrres}, we examine the accuracy of ChatGPT3.5 vs ChatGPT4 where question type and image presence are predictors in an ordinal logistic regression model. 

\section{Results}\label{sec:results}

This investigation began with curiosity about the performance of ChatGPT3.5 and ChatGPT4 on an exam given to seven first-year graduate students in a statistical methods course in the fall of 2022. This is the same ``homemade'' graduate exam mentioned in Section \ref{sec:lit}. Exam questions were loaded one-by-one into either ChatGPT3.5 and ChatGPT4, and the answers from each generative AI platform was graded with the same rubric as the student exams were graded. Grades were collected from the students' answers and from the two versions of ChatGPT.  

\begin{figure}[h]
\centering
\includegraphics[scale=.4]{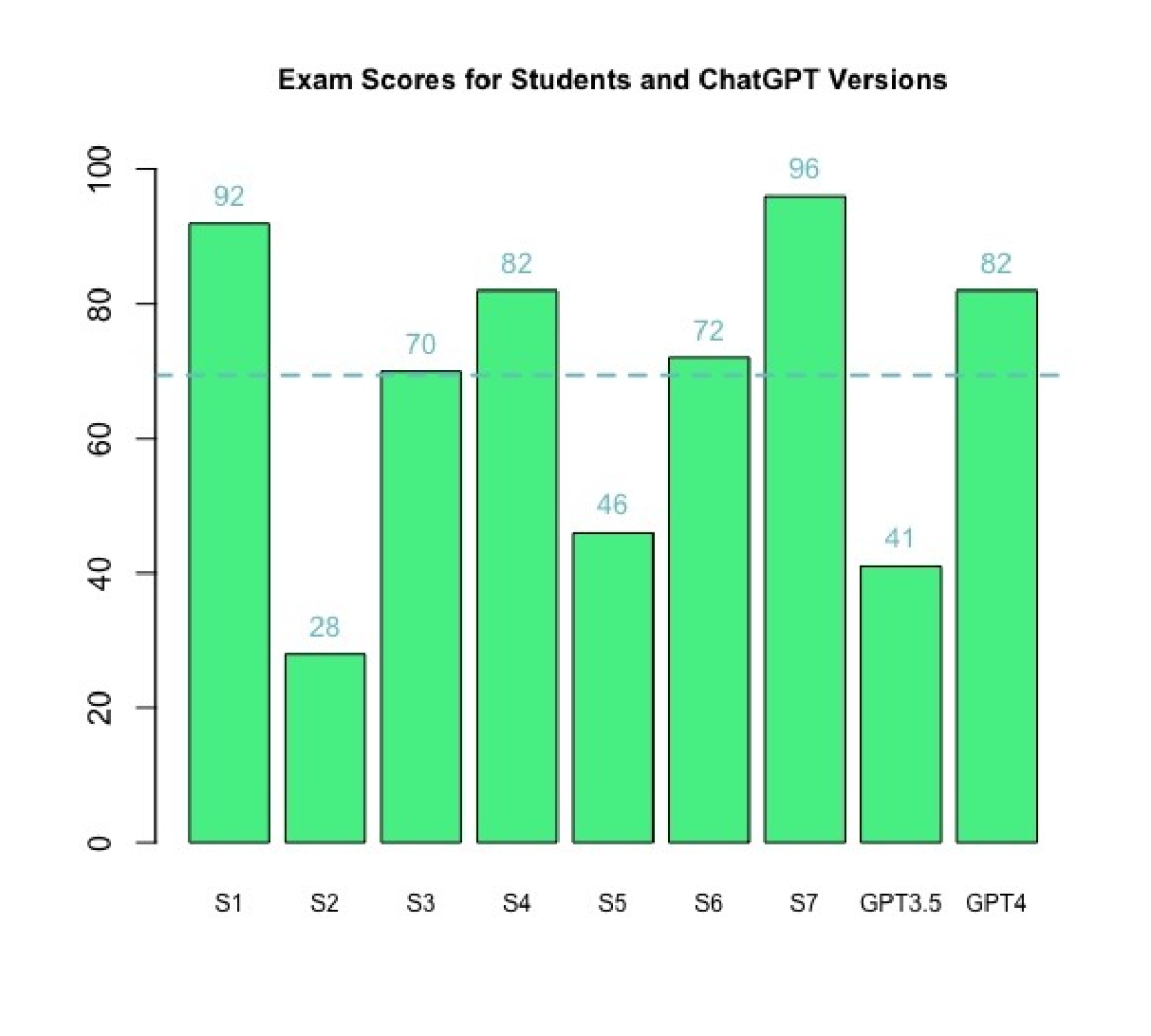}
\caption{Scores of students on the methods exam compared to scores from ChatGPT3.5 and ChatGPT4. Each bar represents a student, and the two bars on the far right represent the generative AI platforms. The dashed horizontal line represents the mean of the students' scores excluding scores from generative AI software}\label{fig:scores}
\end{figure}

Fig \ref{fig:scores} displays the scores for the seven graduate students and the two versions of ChatGPT. Each bar represents a different student or AI platform. The numbers at the top of each bar give the scores, and the dashed horizontal line is the mean of the students' scores, excluding scores from ChatGPT3.5 and ChatGPT4. We see that ChatGPT3.5 has the second worst score on the exam, while ChatGPT4 is tied for the third highest score. Interestingly, if ChatGPT3.5 was a student, it would have failed the exam, while ChatGPT4 would have easily passed. 

We then extended the investigation to include the overall performance of ChatGPT3.5 and ChatGPT4 on each of the nationally normed exams (CAOS, AP, and ACTM). Table \ref{tab:ExamTable} displays the overall percentage correct for each exam by each platform. The overall scores for the first-year exam shown in Fig \ref{fig:scores} are given in the last row for comparison. While each exam has its own official purpose and scale, if we use the traditional passing threshold of 70\% correct, we see that ChatGPT3.5 would have failed all tests while ChatGPT4 would have passed these same tests. Note that if ChatGPT3.5 and ChatGPT4 provided solutions with the same quality, the probability of getting this result by chance is only 6.25\% (1/16). 

\begin{table}[h]
\centering
\caption{Exam Scores from ChatGPT3.5 and ChatGPT4 for each exam used in this study}
\label{tab:ExamTable}
\begin{tabular}{|c|c|c|}
\hline
\textbf{Exam} & \textbf{Score from ChatGPT3.5} & \textbf{Score from ChatGPT4} \\
\hline
ACTM & 64\% & 100\% \\
\hline
AP2011 & 50\% & 81\% \\
\hline
CAOS & 48\% & 70\% \\
\hline
First Year Exam & 41\% & 82\% \\
\hline
\end{tabular}
\end{table}

While the results in Table \ref{tab:ExamTable} are striking, it is important to make an appropriate statistical comparison of the results based on the paired nature of the data. We can use McNemar's test, as explained in Section \ref{sec:mcnemar}, to determine whether the probability of ChatGPT4 getting a question correct and ChatGPT3.5 getting it wrong is equal to the probability of ChatGPT4 getting a question incorrect and ChatGPT3.5 getting a question correct.

Table \ref{tab:mcnemar} shows the number of concordant and discordant questions for each GPT platform for all 93 questions represented on all 4 tests. For example, there were 41 questions of the 93 that ChatGPT4 and ChatGPT3.5 both answered correctly and 11 that they both answered incorrectly.  Additionally, Table \ref{tab:mcnemar} shows that there were 35 questions that ChatGPT4 answered correctly that ChatGPT3.5 did not and only 6 that ChatGPT3.5 answered correctly that ChatGPT4 did not.  The latter are the \textit{discordant} pairs, which is exactly the parameter of interest for McNemar's test. 
\begin{table}[h]
\centering
\caption{Contingency Table for McNemar's Test}
\label{tab:mcnemar}
\begin{tabular}{lcc}
\toprule
 & \textbf{ChatGPT4 Correct} & \textbf{ChatGPT4 Incorrect} \\
\midrule
\textbf{ChatGPT3.5 Correct}   & 41 & 6 \\
\textbf{ChatGPT3.5 Incorrect} & 35 & 11 \\
\bottomrule
\end{tabular}
\end{table}

\begin{figure}[h]
    \centering
    \begin{minipage}{0.48\linewidth}
         \includegraphics[width=0.8\linewidth]{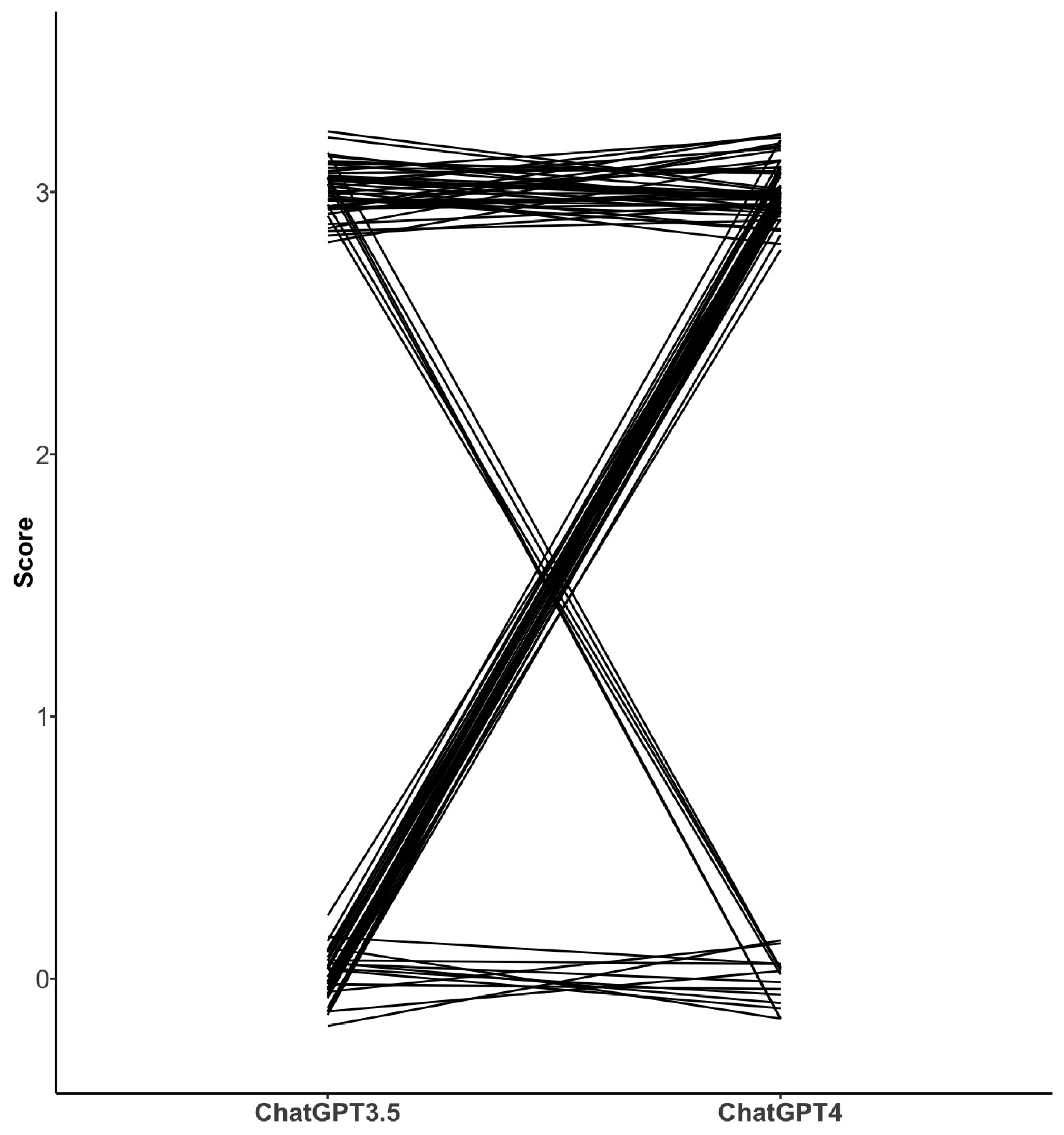}
        \caption{Spaghetti plot showing concordant and discordant pairs. The lines have been jittered for visibility.}
        \label{fig:con-and-disCandI}
    \end{minipage}
    \hfill 
    \begin{minipage}{0.45\linewidth}
        \includegraphics[width=0.8\linewidth]{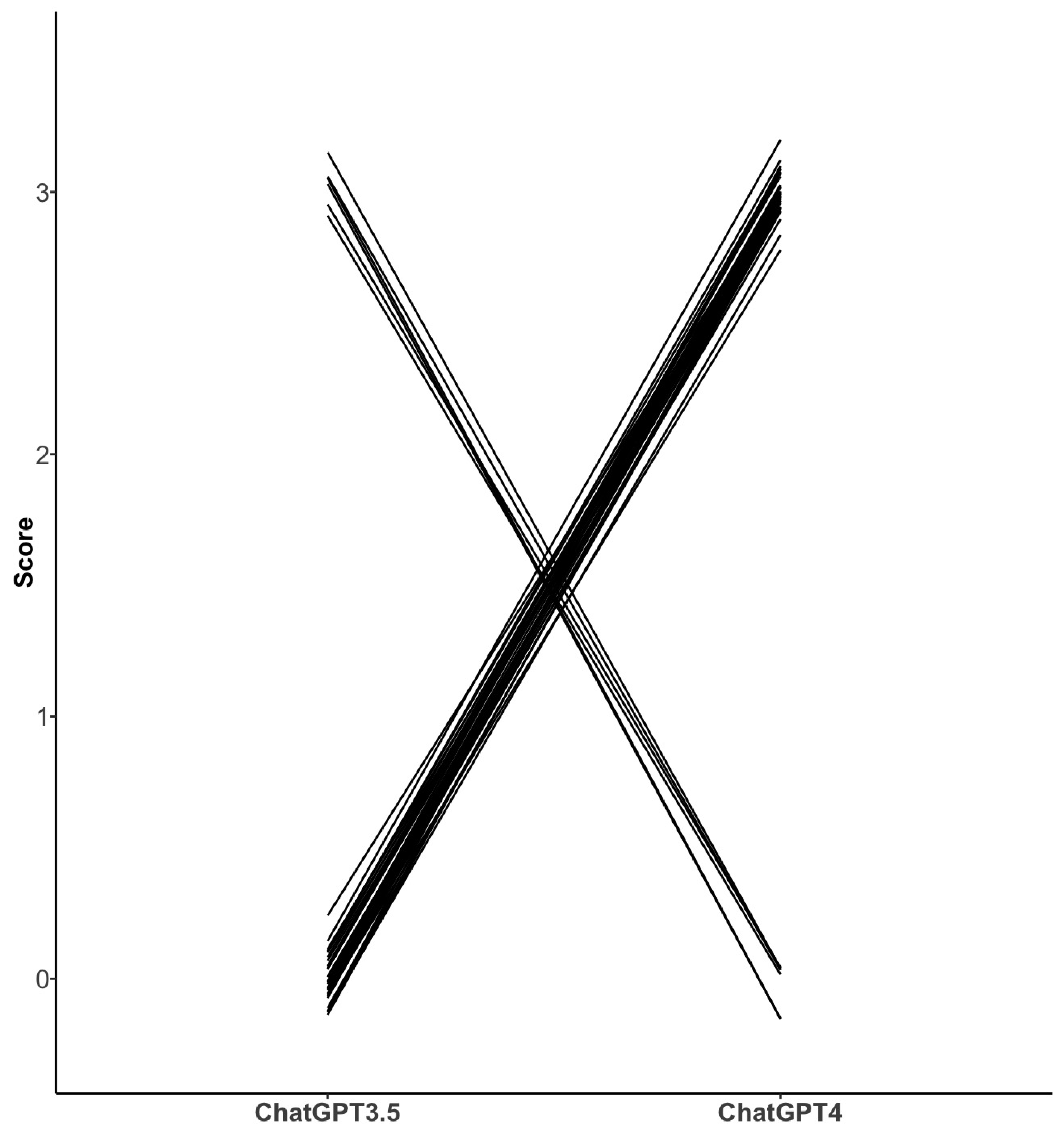}
        \caption{Spaghetti plot showing only discordant pairs. The lines have been jittered for visibility.}
        \label{fig:onlydiscordantCandI}
    \end{minipage}
\end{figure}

These results are represented graphically in the spaghetti plot in Figs \ref{fig:con-and-disCandI} and \ref{fig:onlydiscordantCandI}. Small random numbers have been added to each score (0 and 3) which has the effect of ``jittering'' the lines to prevent them from hiding one another. Without jittering, because all scores are either 0 (incorrect) or 3 (correct), the plot would show single lines from 0 to 0, 0 to 3, 3 to 3, or 3 to 0.   

In Fig \ref{fig:con-and-disCandI}, the flat lines with slope = 0 represent the \textit{concordant} pairs in which both ChatGPT3.5 and ChatGPT4 both had similar quality of answer (both correct or both incorrect).  In Fig \ref{fig:onlydiscordantCandI}, the lines with positive slope reflect the 35 questions that ChatGPT3.5 answered incorrectly but ChatGPT4 answered correctly and the fewer lines with negative slope reflect the 6 questions that ChatGPT3.5 answered correctly but ChatGPT4 did not. Again, the difference in the number of discordant pairs for each platform, highlighted in Fig \ref{fig:onlydiscordantCandI}, are the focus of McNemar's test. 

If there is sufficient evidence to reject $H_0$, then there is sufficient evidence of a difference in performance of ChatGPT3.5 and ChatGPT4.  For these data, McNemar's test yields a p-value of $.000012$ ($\chi^2_1 = 19.9$) which implies that there is a 12 in 10 million chance of seeing a test statistic this large or larger if the null hypothesis is true. This p-value is overwhelming evidence that of the proportion of discordant pairs that ChatGPT4 answered correctly is greater than that of ChatGPT3.5.  

While not the direct metric tested by McNemar's test, it is useful to note that ChatGPT4 answered $80\%\ (76/93)$ of the questions in this study correctly while ChatGPT3.5 only answered $50\%\ (47/93)$ correctly. Therefore, assuming these questions are representative of the types of questions that would be asked on a typical introductory statistics exam and that that a passing grade was 70\%, ChatGPT4 would pass such an exam, and ChatGPT3.5 would not. 

In the previous analysis, we did not consider the effect the type of question. The discrepancy in ChatGPT3.5 scores versus ChatGPT4 scores for the exam represented in Fig \ref{fig:scores} is that ChatGPT3.5 cannot read images or tables. For questions using visualizations, ChatGPT3.5 gave hints for reading the visualizations. For example, when asked to compare the medians of two boxplots, it explained where to find the median of a boxplot in general terms. 

In Section \ref{sec:images}, we perform a formal analysis to consider differences in concordant and discordant pairs conditioned on the need to analyze an image in order to answer multiple choice questions. In section \ref{sec:free}, we calculate McNemar's test statistic for free response and multiple choice questions separately. Then, in section \ref{sec:olrres}, we examine ordinal logistic regression models to determine the relationship between the score given to a free response question based on Table \ref{tab:rubric}.

\subsection{Examining Questions with and without Images}\label{sec:images}

As noted in Table \ref{tab:comparison}, of the 63 questions that did not have images, there were 17 questions that ChatGPT4 answered correctly that ChatGPT3.5 did not while there were only 6 that ChatGPT3.5 answered correctly that ChatGPT4 did not.  McNemar's test yields a p-value of .03 for these data indicating a 3 in 100 chance of seeing a test statistic as large or larger than the one calculated if the null hypothesis is true. This p-value is sufficient evidence of a difference in performance of ChatGPT3.5 and ChatGPT4.  This difference is reflected in the fact that ChatGPT4 answered $88\%$ of questions without images correctly while ChatGPT3.5 only answered $71\%$ of the same questions correctly.  

\begin{table}[hb]
\centering
\caption{Contingency Table for McNemar's test for multiple choice questions without use of an image as part of the question.}
\label{tab:comparison}
\begin{tabular}{lcc}
\toprule
 & \textbf{ChatGPT4 Correct} & \textbf{ChatGPT4 Incorrect} \\
\midrule
\textbf{ChatGPT3.5 Correct}   & 39 & 6 \\
\textbf{ChatGPT3.5 Incorrect} & 17 & 1 \\
\bottomrule
\end{tabular}
\end{table}

As noted in Table \ref{tab:mcnemar_analysis}, of the 30 questions that included images, there were 18 that ChatGPT4 answered correctly that ChatGPT3.5 did not and \textbf{none} that ChatGPT3.5 answered correctly that ChatGPT4 did not. In fact, when an image was presented to ChatGPT3.5 as part of a question, ChatGPT3.5 often responded with ``ChatGPT3.5 cannot directly read or interpret images. It's a text-based model and can only process and generate text.'' McNemar's test yields a p-value of .00006 for these data indicating overwhelming evidence of a difference in performance of ChatGPT3.5 and ChatGPT4.  This difference is reflected in the fact that ChatGPT4 answered $66\%$ (20/30) of questions with images correctly while ChatGPT3.5 only answered $6\%$ (2/30) of the same questions correctly.

\begin{table}[ht]
\centering
\caption{Contingency Table for McNemar's Test Analysis for multiple choice questions referencing an image as part of the question.}
\label{tab:mcnemar_analysis}
\begin{tabular}{lcc}
\toprule
 & \textbf{ChatGPT4 Correct} & \textbf{ChatGPT4 Incorrect} \\
\midrule
\textbf{ChatGPT3.5 Correct}   & 2 & 0 \\
\textbf{ChatGPT3.5 Incorrect} & 18 & 10 \\
\bottomrule
\end{tabular}
\end{table}

\subsection{Analysis by Question Type}\label{sec:free}

In this section we analyze correct and incorrect results separately for free response questions and multiple choice questions. Answers to free response questions were dichotomized into ``correct'' if the score on a question was $\geq 3$ as given by Table \ref{tab:rubric} and ``incorrect'' if the score was $< 3$. As noted in Table \ref{tab:method_comparison}, of the 28 free response questions, there were 18 that ChatGPT4 answered correctly that ChatGPT3.5 did not and only 2 that ChatGPT3.5 answered correctly that ChatGPT4 did not.  McNemar's test yields a p-value of .010 for these data indicating sufficient evidence of a difference in performance of ChatGPT3.5 and ChatGPT4 for free response questions.  This difference is reflected in the fact that ChatGPT4 answered $82\%$ (23/28) of free response correctly while ChatGPT 3.5 only answered $43\%$ (12/28) of the same questions correctly.

\begin{table}[h]
\centering
\caption{Contingency Table for ChatGPT Comparison}
\label{tab:method_comparison}
\begin{tabular}{lcc}
\toprule
 & \textbf{ChatGPT4 Correct} & \textbf{ChatGPT4 Incorrect} \\
\midrule
\textbf{ChatGPT3.5 Correct}   & 10 & 2 \\
\textbf{ChatGPT3.5 Incorrect} & 13 & 3 \\
\bottomrule
\end{tabular}
\end{table}

As noted in Table \ref{tab:mcnemar_new}, of the 65 multiple choice questions, there were 22 that ChatGPT 4 answered correctly that ChatGPT 3.5 did not and only 4 that ChatGPT 3.5 answered correctly that ChatGPT 4 did not.  McNemar's test yields a p-value of .0009 for these data indicating strong evidence of a difference in performance of ChatGPT 3.5 and 4 for multiple choice questions.  This difference is reflected in the fact that ChatGPT 4 answered $81\%$ (23/28) of multiple choice questions correctly while ChatGPT 3.5 only answered $54\%$ (12/28) of the same questions correctly.

\begin{table}[h]
\centering
\caption{Contingency Table for McNemar's Test Between Two Methods}
\label{tab:mcnemar_new}
\begin{tabular}{lcc}
\toprule
 & \textbf{ChatGPT4 Correct} & \textbf{ChatGPT4 Incorrect} \\
\midrule
\textbf{ChatGPT3.5 Correct}   & 31 & 4 \\
\textbf{ChatGPT3.5 Incorrect} & 22 & 8 \\
\bottomrule
\end{tabular}
\end{table}

\subsection{Ordinal Logistic Regression Results}\label{sec:olrres}

Figs \ref{fig:con-and-disCandI} and \ref{fig:onlydiscordantCandI} showed answers from ChatGPT3.5 and ChatGPT4 graded as either ``correct'' or ``incorrect''. In other words, the focus was on the accuracy of the answer to each questions. In this section, we focus on the quality of the answer between ChatGPT3.5 to ChatGPT4 for through the scale for free response questions introduced in Table \ref{tab:rubric}. 

\begin{figure}[h]
    \centering
    \begin{minipage}{0.45\linewidth}
        \includegraphics[width=\linewidth]{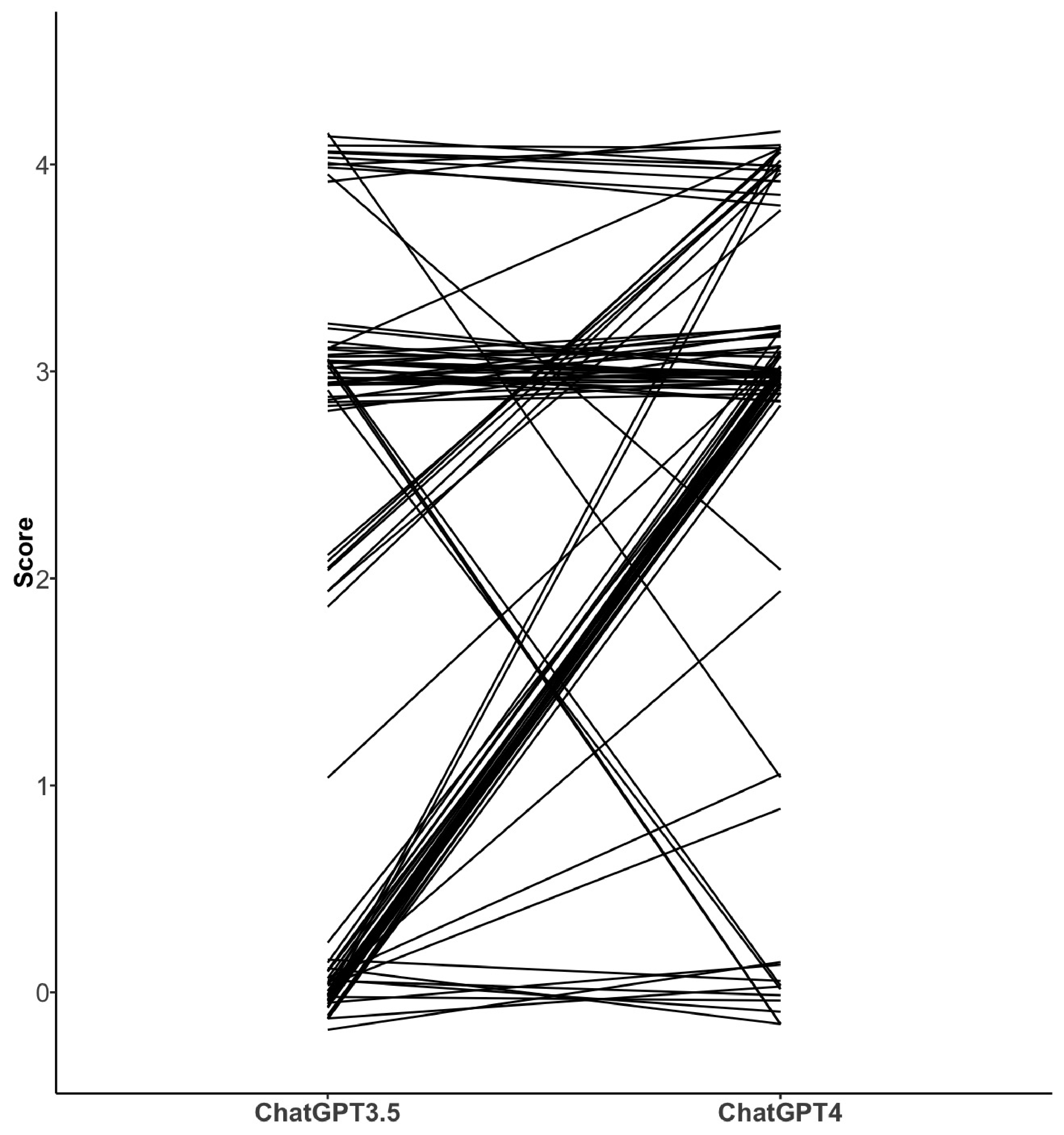}
        \caption{Spaghetti plot showing concordant and discordant pairs. The lines have been jittered for visibility.}
        \label{fig:con-and-dis}
    \end{minipage}
    \hfill 
    \begin{minipage}{0.45\linewidth}
        \includegraphics[width=\linewidth]{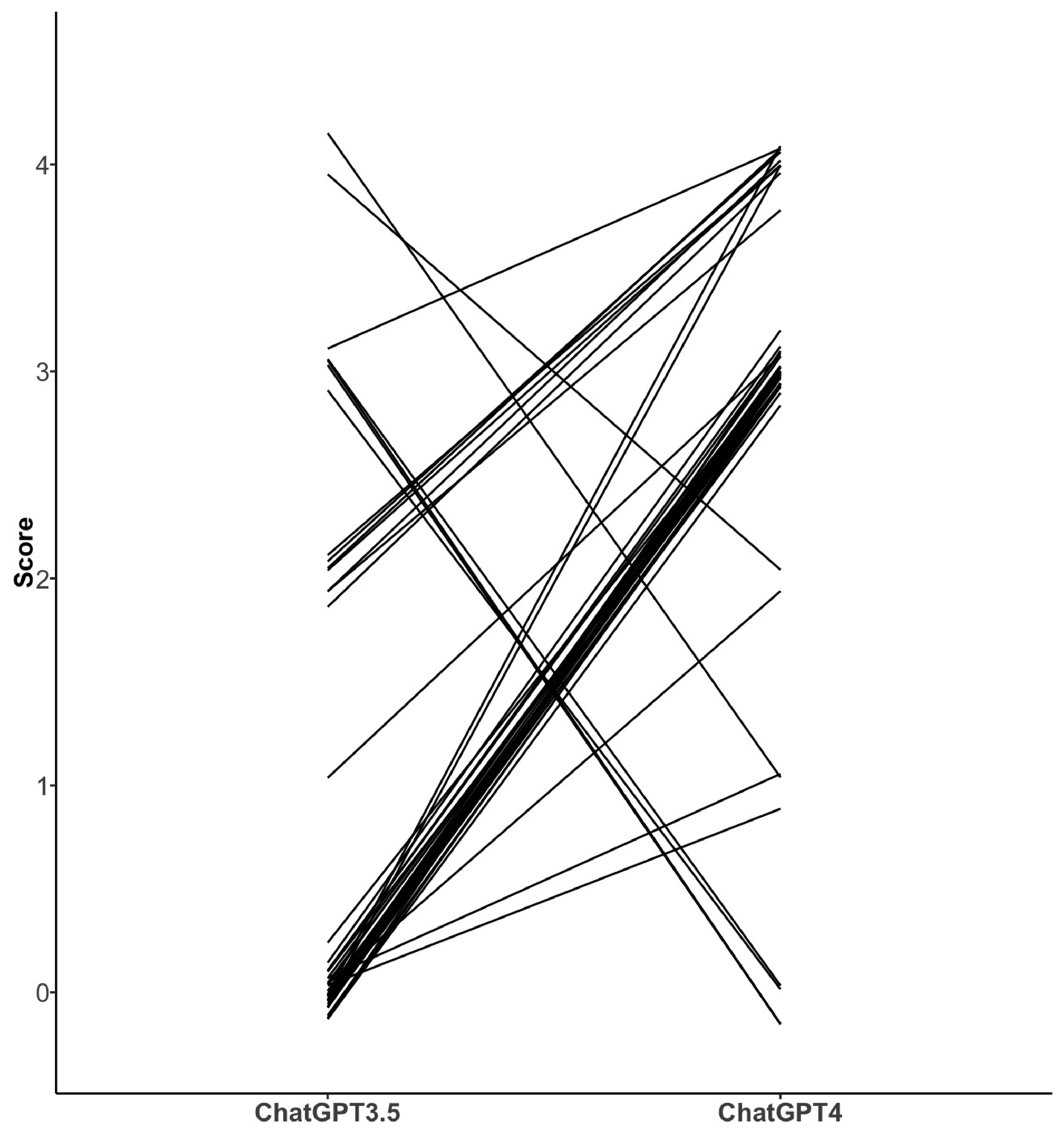}
       \caption{Spaghetti plot showing only discordant pairs. The lines have been jittered for visibility.} \label{fig:no-discordant}
    \end{minipage}
\end{figure}

Fig \ref{fig:con-and-dis} illustrates this difference in quality score from ChatGPT3.5 to ChatGPT4 for all 93 questions.  The scores have been jittered to prevent lines from hiding one another. The flat lines represent the 52 questions in which ChatGPT3.5 and ChatGPT4 had the same quality (the concordant pairs). The positively sloped lines indicate the 35 questions where ChatGPT4 provided a higher quality response while negatively sloped lines indicate 6 questions in which ChatGPT3.5 provided a higher quality answer. Fig \ref{fig:no-discordant} displays the same information but removes the concordant pairs to highlight the questions that reflect the difference in quality (the discordant pairs).  

To determine the effect of type of question (free response or multiple choice) and presence of an image within the question (Image or Not), we performed an ordinal logistic regression where the response was given on a scale of 0 to 4 as shown in Table \ref{tab:rubric}. The parameter estimates and associated statistics for each coefficient in the model are shown in Table \ref{tab:ParamEstTable}. 

\begin{table}[hb]
\centering
\caption{Model Coefficients from Ordinal Logistic Regression}
\label{tab:ParamEstTable}
\begin{tabular}{lcccc}
\toprule
\textbf{Coefficient} & \textbf{Estimate} & \textbf{Std. Error} & \textbf{z value} & \textbf{Pr$(>|z|)$} \\
\midrule
(Intercept):Much & -1.8256 & 0.3602 & -5.068 & $4.02 \times 10^{-7}$ *** \\
(Intercept):Somewhat & -1.1350 & 0.3237 & -3.506 & $4.55 \times 10^{-4}$ *** \\
(Intercept):Same & 2.2053  & 0.4466 & 4.938  & $7.88 \times 10^{-7}$ *** \\
TypeFR           & 0.4565  & 0.4538 & 1.006  & 0.314439 \\
Image Present           & 1.9172  & 0.4599 & 4.169  & $3.06 \times 10^{-5}$ *** \\
\bottomrule
\end{tabular}
\end{table}

In the far left column of Table \ref{tab:ParamEstTable}, we have a name for each coefficient in the model. Starting from the bottom up, the row entitled ``Image Present'' in the table shows the coefficient and associated statistics for whether reference to an image is necessary to answer the question. For any given question type, the odds of ChatGPT4 providing a higher quality response than ChatGPT3.5 for questions with an image are estimated to be $580\%$ ($e^{1.917} - 1$) more than that of questions without an image. We are $95\%$ confident that this increase in odds is between $276\%$ and $1575\%$, and the p-value for this coefficient is 0 to four decimal places.  In short, the evidence strongly suggests that the odds of ChatGPT4 having an advantage over ChatGPT3.5 increase enormously when the question has an image compared to those questions that do not, regardless of the the question type.  This makes sense as ChaptGPT3.5 is unable to digest image data.  

The penultimate row of Table \ref{tab:ParamEstTable} shows the value of the coefficient for type of question, where the reference category is multiple choice. For a fixed image status, the odds of ChatGPT4 having a higher quality response than ChatGPT3.5 for a free response question is estimated to be 57\% ($e^{.4565}-1$) greater than the same odds for multiple choice questions. A $95\%$ confidence interval for this percent change in odds is between a $35.2\%$ decrease and a $284\%$ increase, and the p-value for this predictor is approximately $.31$. There is not enough evidence to suggest an effect of the question type on the odds of ChatGPT4 performing better than ChatGPT3.5 after controlling for the presence of an image in the question. Even though 100\% (even odds) is within the confidence interval, most of the estimated odds within the interval are much larger than 100\%, indicating that there is some evidence that ChatGPT4 is more likely to perform better than ChatGPT3.5 if the question is a free response  question rather than a multiple choice question.

The top three rows of the parameter estimate table are comprised of the intercepts which are used to obtain estimated probabilities of ChatGPT4 to provide a higher quality response than ChatGPT3.5 for the four different types of questions. For instance for free response questions without an image, the estimated probability of ChatGPT4 providing a response that is much better than ChatGPT3.5 is (see Eq. \ref{eq:prob})

\begin{align}
P(Y = \text{Much} | \text{FR} = 1, \text{Image} = 1) &= \frac{e^{\alpha_{\text{Much}} + \beta_{\text{FR}} \cdot \text{FR} + \beta_{\text{Image}} \cdot \text{Image}}}{1 + e^{\alpha_{\text{Much}} + \beta_{\text{FR}} \cdot \text{FR} + \beta_{\text{Image}} \cdot \text{Image}}} \nonumber\\
 &= \frac{e^{-1.8256+.4565 + 1.9172}}{1 + e^{-1.8256+.4565 + 1.9172}}\\  &= .6337 \nonumber
\end{align}

Therefore, the estimated probability of ChatGPT4 providing a much higher quality response than ChatGPT3.5 for free response questions without images is .6337 which is striking but not surprising since ChatGPT3.5 cannot process image data.   

Similarly from Eq. \ref{eq:prob}, $P(Y = \text{Somewhat} | \text{FR} = 1, \text{Image} = 1) $
\begin{align}
&=& 
\frac{e^{\alpha_{\text{Somewhat}} + \beta_{\text{FR}} \cdot \text{FR}+ \beta_{\text{Image}} \cdot \text{Image}}}{1 + e^{\alpha_{\text{Somewhat}} + \beta_{\text{FR}} \cdot \text{FR}+ \beta_{\text{Image}} \cdot \text{Image}}} -  \frac{e^{\alpha_{\text{Much}} + \beta_{\text{FR}} \cdot \text{FR}+ \beta_{\text{Image}} \cdot \text{Image}}}{1 + e^{\alpha_{\text{Much}} + \beta_{\text{FR}} \cdot \text{FR}+ \beta_{\text{Image}} \cdot \text{Image}}}
\\ \nonumber
& = &\frac{e^{-1.1350+.4565+ 1.9172}}{1 + e^{-1.1350+.4565+ 1.9172}} - \frac{e^{-1.8256+.4565+ 1.9172}}{1 + e^{-1.8256+.4565+ 1.9172}} = .1417
\end{align}

Therefore, the estimated probability of ChatGPT4 providing a somewhat higher quality response than ChatGPT3.5 for free response questions without images is .1417. Given these two probabilities, we can estimate that for free response questions with images, ChatGPT4 will provide a higher quality response with probability .6337 + .1417 = .7754 and is thus more likely than not to provide a higher quality response than ChatGPT3.5 for these questions. 

Similar calculations for each type of question yields Table \ref{tab:ProbabilityTable} in which the rows represent the probability of a particular ChatGPT4 performance with respect to ChatGPT3.5 and the columns represent each of the four types of questions.  The top two rows, corresponding to Much and Somewhat, indicate probabilities that ChatGPT4 will provide higher quality responses than ChatGPT3.5 while the third row indicates the probabilities that ChatGPT4 and ChatGPT3.5 will provide the same quality response and the bottom row indicates the probability that ChatGPT3.5 will actually provide a higher quality response than ChatGPT4.  

Non zero sums of the top two rows for any question type indicate an advantage for those using ChatGPT4.  \textbf{All} of these sums are greater than 20\% with the entries highlighted in green in Table \ref{tab:ProbabilityTable} summing to greater than 50\% indicating that for those question types, ChatGPT4 is more likely than not to provide a higher quality response. 

\begin{table}[h]
\centering
\caption{Probabilities of responses across different question types and image conditions}
\label{tab:ProbabilityTable}
\begin{tabular}{@{}lcccc@{}}
\toprule
\textbf{Response Type} & \textbf{FR w/o Image} & \textbf{FR w/ Image} & \textbf{MC w/o Image} & \textbf{MC w/ Image} \\ \midrule
\textbf{P(Much)}      & 0.2028 & \cellcolor{green!20}0.6337 & 0.1388 & \cellcolor{green!20}0.5229 \\
\textbf{P(Somewhat)}  & 0.1338 & \cellcolor{green!20}0.1417 & 0.1045 & \cellcolor{green!20}0.1633 \\
\textbf{P(Same)}      & 0.5981 & 0.2145 & 0.6575 & 0.2979 \\
\textbf{P(Worse)}     & 0.0653 & 0.0102 & 0.0993 & 0.0159 \\
\bottomrule
\end{tabular}
\label{tab:response_probabilities}
\end{table}

Clearly, by all measures, ChatGPT4, which requires a subscription, outperforms ChatGPT3.5, which is free. The difference in performance has implications for the use of generative AI as individual tutors for statistics courses. The platform used will determine the quality of the instruction. Note that, while we  examine the quality of results from the platforms using exams, we do not condone the use of generative AI to take exams. These exams contain questions that are proxies that might be used to examine student learning outcomes in a typical statistics course. If students use generative AI as tutor, a tutor based on ChatGPT4 is more often correct than a tutor based on ChatGPT3.5, which puts students who cannot pay for ChatGPT4 at a disadvantage.

\section{Discussion}

Sal Kahn of Khan Academy has, since 2008, advocated for every child in the World to have a personal tutor.  In that year, he put his educational videos on YouTube in an effort to democratize education for all.  His mission to provide everyone with their own tutor was given a great boon by the emergence of ChatGPT's potential to provide insights into a wide variety of subjects with the knowledge of what it was trained on, nearly the entire Internet.  OpenAI, the company that developed ChatGPT, saw this great potential early and made its ground breaking GPT3 technology available to Dr. Kahn in June of 2022 \cite{khanmigo_article}, six months before its blockbuster release in November 2022.  The eventual product to emerge was named Kahnmigo and was hailed as a great leap forward towards accomplishing his mission of a personal tutor for all. \cite{KhanAcademy2024}

While Sal Kahn's goal was to democratize education and bridge the educational divide between different socioeconomic classes, ironically, the introduction of a more powerful paid version of this technology (ChatGPT4) introduces the potential to increase this divide.  The premise in which we address the potential equity issue posed here is that good tutoring is delivered in two ways: 

\begin{itemize}
    \item[]A. Through good general tips, hints, metaphors and examples
    \item[]B. Through thorough hints and solutions to specific practice problems.
\end{itemize}

Similar to how a human tutor is often assessed, a proxy for these two tenants of tutoring will be measured performance on practice problems in statistics that are similar to those that would be seen in class and that are aimed at facilitating the mastery of the desired skill and methods. In order to detect and measure a difference in quality between ChatGPT3.5 and ChatGPT4 in their responses to potential practice questions, we have partitioned the question bank into those with images and those without images.  

ChaptGPT3.5 is not able to ingest images; thus for questions for which reference to an image is part of the question, ChatGPT3.5 is limited to the text.  This limitation is shown in that, of the 30 questions that had images, both platforms answered 2 of them correctly (upper left in Table \ref{tab:mcnemar_analysis}), both platforms answered 10 of them incorrectly (lower right in Table \ref{tab:mcnemar_analysis}) and ChatGPT4 provided correct responses to the remaining 18 questions (lower left in Table \ref{tab:mcnemar_analysis}) that ChatGPT3.5 subsequently could not provide a response and were thus scored as ``incorrect'' (upper right in Table \ref{tab:mcnemar_analysis}). There is overwhelming evidence to suggest that of the questions that the platforms disagree on (the discordant pairs). ChatGPT4 will be more likely to answer correctly with respect to ChatGPT3.5 (p-value $< .0001$ from McNemar's test).  For reference, in our study ChatGPT4 answered 66\% of questions correctly while ChatGPT3.5 was only able to to provide correct solutions to 6\% of these questions. Furthermore, recall our ordinal logistic regression analysis yielded estimated probabilities of over 60\% that ChaptGPT4 would provide a higher quality response that ChatGPT3.5 for questions with images (free response and multiple choice).  However, we did not examine the effect of alt text, which is text embedded within a visualization to be read aloud by screen readers, contained within plots on the interpretability of images for either generative AI platform.

While questions with images exposed an important difference in quality between the two platforms, there was also evidence of a difference between the platforms for questions that did not have images.  Overall, ChatGPT4 answered 88\% of questions without images correctly while ChatGPT3.5 only answered 71\% of the same questions with a correct response. If translated into grades, ChatGPT4 would receive a B$+$ while ChatGPT3.5 would receive a C$-$. For the discordant pairs, there was again sufficient evidence to suggest that ChatGPT4 is more likely to provide a correct solution (p-value from McNemar's test was .03) and it was estimated that for questions without images there was a greater than 20\% chance that ChatGPT4 would provide a higher quality response than ChatGPT3.5 for questions without images regardless of whether the quesiton was free response or multiple choice.

Interestingly, while ChatGPT4 was found to be more likely to provide higher quality solutions to both questions with and without images, we are 95\% confident that the odds of ChatGPT4 providing higher quality solutions than ChatGPT3.5 for questions with images is between 176\% and 1575\% more than the odds of a higher quality response for questions without images (p-value = .00007 from an ordinal logistic regression).  

It is also intuitive that platforms may have different advantages when it comes to the modality of the question: multiple choice versus free response.  For instance, it may be that both platforms perform similarly well on multiple choice questions and that equity concerns are minimized for these types of questions.  The evidence strongly suggested that ChatGPT4 provided higher quality responses (increased likelihood of being correct) for both multiple choice (McNemar p-value = .0009) and free response questions (McNemar p-value = .010).  Furthermore, the estimated probability of ChatGPT4 providing a higher quality response than ChatGPT3.5 was at least 23\% for free response questions and at least 33\% for multiple choice questions.  

A limitation of this study is that the sample size was not large enough to statistically exclude the plausibility that the difference detected in the modality of the question is confounded with the presence of an image; however, the empirical evidence is compelling.  Of the eight free response questions that had images, ChatGPT4 provided a correct solution for five (62.5\%) while ChatGPT3.5 did not answer any correctly.  Additionally, of the remaining 20 free response questions without images, ChatGPT4 answered 18 (90\%) correctly while ChaptGPT3.5 was only able to answer ten (50\%) correctly.  Furthermore, ChatGPT4 answered 80\% of the the questions that made up the 10 discordant pairs correctly.

Another potential confounding variable in the study is the difficulty of the question; a proxy for this is the exam to which each question belongs.  The 93 questions in this study were sourced from four different tests of variable difficulty.  The most introductory was the ACTM as it focused on assessing high school statistical skill.  The next level up was the CAOS exam which is focused on introductory college level statistics student, while the AP exam is aimed at more advanced students who are looking to earn credit for having completed a introductory college statistics course.  The span of difficulty was capped at the graduate statistics level from one of the author's graduate statistics exams.  Again, while this study did not have enough power to statistically test for these effects, the empirical evidence was very convincing as ChatGPT4 outperformed ChatGPT3.5 on every exam (see Table \ref{tab:ExamTable}). 

To this point, this study has provided evidence to suggest that ChaptGPT4 is more likely to provide higher quality responses to questions regardless of their image content and modality. We submit that it is reasonable to conclude that those that must use ChatGPT3.5 are at a greater chance of receiving poorer quality, and often incorrect, responses than those that use ChatGPT4.  Given that ChatGPT4 is currently \$20 a month, it stands to reason a significant equity concern exists between those who can afford, or choose to purchase, ChatGPT4 and those who cannot, or choose not to.  To this last point, this equity concern transcends the socioeconomic concerns as those that simply do not want to put their financial information (ie. credit card number) at risk may also be forced to use the free ChatGPT3.5.  

Looking forward, it is reasonable to assume that as soon as the publication date of this study new versions of ChatGPT will have become available.  While similar studies will need to be conducted to monitor the presence and extent of equity concerns, the performance of ChatGPT4 in this study suggests that equity concerns are likely to persist into at least the near future.  In this study, ChatGPT4 answered 18.2\% (17/93) of the questions incorrectly and was even outperformed by ChatGPT3.5 on six of the 93 (see Table \ref{tab:mcnemar}).  Therefore, even if ChatGPT4 was soon made available for free, those that had access to more advanced ChatGPTs would have, an estimated, up to 18\% increase in likelihood of receiving a quality / correct response.  

\section{Extensions and Future Work}

On May 13, 2024, Open AI announced the release of ChatGPT4o. According to OpenAI, the release of ChatGPT4o will make more features of ChatGPT4 available to without a (paid) subscription.  These features include access to ``GPT-4 level intelligence'' and the ability to read and upload images \cite{OpenAI2024}. However, they also mentioned,

\begin{quote}
There will be a limit on the number of messages that free users can send with GPT-4o depending on usage and demand. When the limit is reached, ChatGPT will automatically switch to GPT-3.5 so users can continue their conversations. \cite{OpenAI2024}
\end{quote}

As of the submission of this paper, this limit has yet to be defined; however, since paid users will not experience usage limits, the existence of the limit means the persistence of the equity concerns described in this study. Therefore, the most immediate extension of this paper will be to assess just how much better ChatGPT4o is than ChatGPT4 at ``understanding'' images with respect to statistics questions. Currently, the performance of ChatGPT4o is quite impressive. Users can hold a smart phone above a paper on which there is a hand-written question, and ChatGPT4o can read the equation and give hints on how to solve it. This remarkable ability was demonstrated in a video using a simple algebraic question with one variable \cite{OpenAI2024}. It is unclear how ChatGPT4o would respond to a system of hand-written equations, or a complicated multiple integral, or in determining differences in medians among a set of hand-drawn boxplots.  

In terms of equity for visually impaired individuals, the video demonstration of ChatGPT4o showed remarkable text to speech capability, both in ChatGPT4o's ability  to understand speech and to respond in clear speech. ChatGPT4o can vary pitch and tone to be more dramatic or to speak like a robot. Therefore, it seems that visually impaired individuals can use ChatGPT4o with little trouble. However, during the demonstration of ChatGPT4o's ability to read handwritten equations on paper, the demonstrator held his phone above the page, which might not be possible for someone with limited arm mobility.  Clearly, there are some access issues still to be resolved.

In addition to testing ChatGPT4 vs. ChatGPT4o, the research can be extended to other generative AI programs that are used in classroom teaching. For example, many universities have a site-license for Office365, which includes use of Microsoft Co-Pilot. Other platforms, such as Google Gemini, Anthropic's Claude series of foundational models, and free models from Hugging Face \cite{wolf2020huggingfaces} would also be useful to explore and could reduce or eliminate the equity concerns identified above. Some preliminary work comparing ChatGPT3.5, ChatGPT4, Claude, Gemini, Pi, and CoPilot, on a different set of questions from statistics exams than used in this research showed that ChatGPT4, the only subscription version, is the clear winner in terms of accuracy. The other platforms perform approximately the same \cite{doyle}. 

We would further like to examine data analyses from upper level required and elective courses, analyses of high dimensional data sets, and data analysis projects assigned in other STEM disciplines. We are further open to examining ways to automate the collection, analysis, and reporting of the responses to prompts in ChatGPT in order to facilitate comparison of a much broader collection of questions and answers. In particular, we did not ask ChatGPT3.5 or ChatGPT4 to write or correct code for a data analysis, which is something that generative AI platforms are purported to do quite well \cite{openai2024gpt4}.

Finally, ChatGPT3.5 and ChatGPT4 give responses to questions that could contain socioeconomic or cultural bias due to word choice of the input. It is important to perform future research on how humans (particularly students) interpret the answer that the model gives. For example, some answers will have different meanings to people from different countries or cultures. Different meanings based on cultural backgrounds might not be as much of an issue in quantitative disciplines such as statistics, but there might be a clear impact in the humanities and social sciences.

\backmatter

\begin{appendices}

\section{Samples of the Four Types of Questions Entered into ChatGPT3.5 and ChatGPT4}\label{secA1}

\begin{figure}[htb]
    \centering
    \includegraphics[scale=.9]{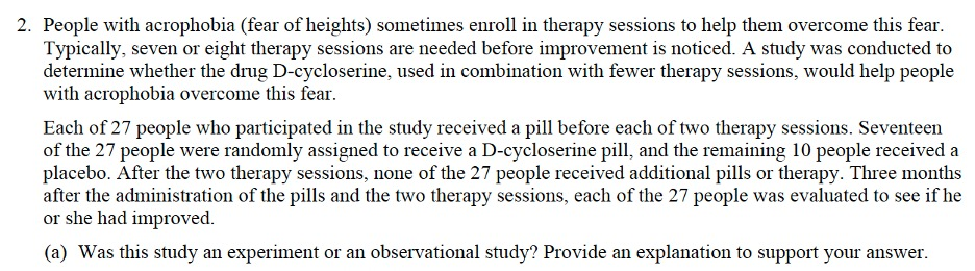}
    \caption{An example of a free response question. The user must enter an answer in sentence form or calculate the answer given information in the problem.}
    \label{fig:fr}
\end{figure}

\begin{figure}[htb]
    \centering
    \includegraphics[scale=1.2]{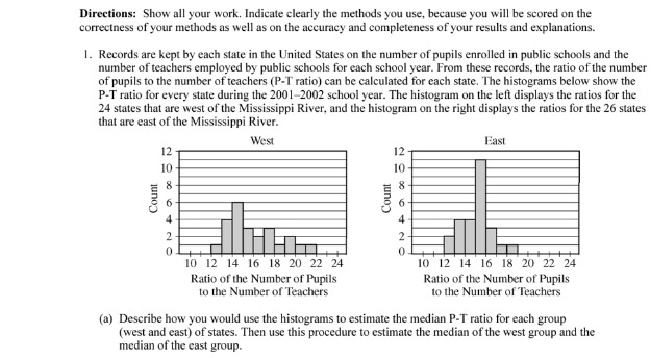}
    \caption{An example of a free response question referencing an image. The user must use information from the image to answer the question.}
    \label{fig:fri}
\end{figure}

\begin{figure}[htb]
    \centering
    \includegraphics[scale=.8]{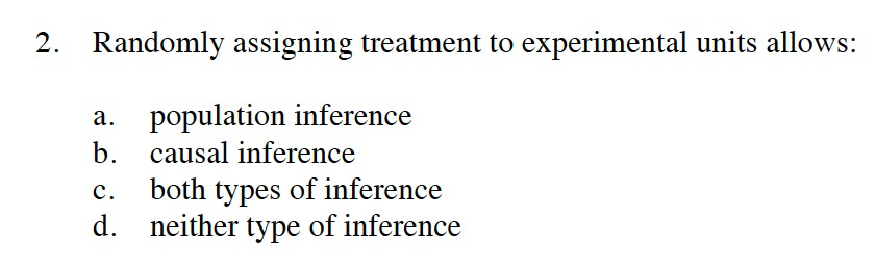}
    \caption{An example of a multiple choice question with responses given.}
    \label{fig:mc}
\end{figure}

\begin{figure}[htb]
    \centering    \includegraphics[scale=.7]{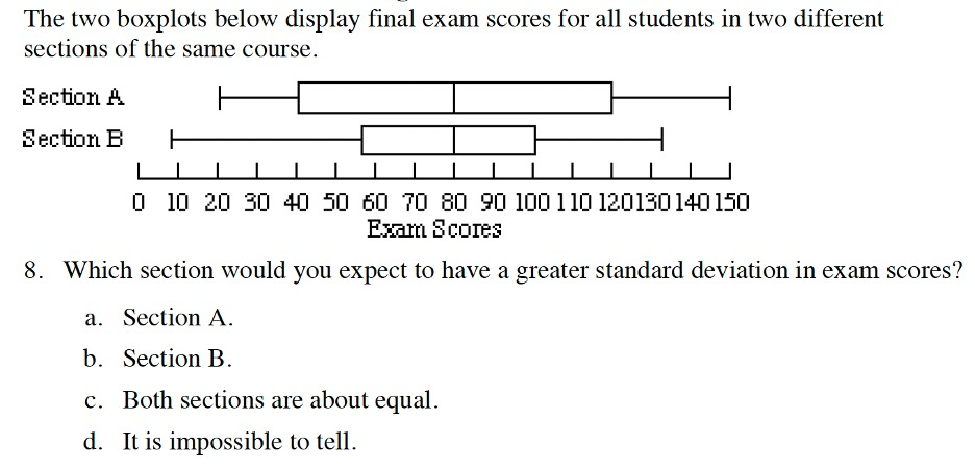}
    \caption{An example of a multiple choice question where the user must examine an image to answer the question.}
    \label{fig:mci}
\end{figure}

In this Appendix, we provide examples of the four types of questions examined in this research.  Fig \ref{fig:fr} is representative of a free response question, and fig \ref{fig:fri} shows a free response question where the user must interpret an image. Fig \ref{fig:mc} shows a question representative of a multiple choice question. Fig \ref{fig:mci} shows a multiple choice question where the student must examine an image in order to answer the question.

\end{appendices}

\clearpage

\bibliography{chatgpt}

\begin{thebibliography}{37}
\bibitem[\protect\citeauthoryear{Terry}{}]{terry}
\begin{botherref}
\oauthor{\bsnm{Terry}, \binits{O.K.}}:
I am a student: You have no idea how much we are using chat{GPT}.
The Chronical of Higher Education
\end{botherref}
\endbibitem
\end{thebibliography}

\end{document}